# The effect of point mutations on energy conduction pathways in proteins

# (The disease mutation A77V in Ryanodine receptor RyR2 induces changes in energy conduction pathways in the protein)

B. Nazan Walpoth
Swiss Cardiovascular Center
University of Bern Inselspital
Cardiology, Bern Switzerland

Burak Erman
Chemical and Biological Engineering, Koc University
Istanbul, Turkey

**Abstract**

Energetically responsive residues of the 173 amino acid N-terminal domain of the cardiac Ryanodine receptor RyR2 are identified by a simple elastic net model. Residues that respond in a correlated way to fluctuations of spatially neighboring residues specify a hydrogen bonded path through the protein. The evolutionarily conserved residues of the protein are all located on this path or in its close proximity. All of the residues of the path are either located on the two Mir domains of the protein or are hydrogen bonded them. Two calcium binding residues, E171 and E173, are proposed as potential binding region, based on insights gained from the elastic net analysis of another calcium channel receptor, the inositol 1,4,5-triphosphate receptor, IP3R. Analysis of the disease causing A77V mutated RyR2 showed that the path is disrupted by the loss of energy responsiveness of certain residues.

**Introduction**: The cardiac Ryanodine receptor has become a subject of increasing interest as its role in the etiology of cardiac disease became more apparent. Two genetic diseases have been linked to mutations in the cardiac Ryanodine receptor: arrhythmogenic right ventricular dysplasia type 2 (ARVD/C Typ 2 or simply ARVD/C) and catecholaminergic polymorphic ventricular tachycardia (CPVT) or familial polymorphic ventricular tachycardia. ARVD/C is an autosomal dominant cardiomyopathy, characterized by partial degeneration of the myocardium of the right ventricle, electrical instability, and sudden death. ARVD/C and CPVT can be caused by mutations in the cardiac Ryanodine receptor 2 gene (RyR2), located on chromosome 1q42.1–q43. There is a familial occurrence in about 50%. ARVD/C is characterized by fibrofatty replacement, primarily of the right ventricle [1]. The gold standard for ARVD/C diagnosis is the demonstration of this alteration of ventricular myocardium, either postmortem or at surgery [2]. In the past 10 years, the identification of causative mutations in plakoglobin, desmoplakin, plakophilin-2, desmoglein-2,and desmocollin-2 has fostered the view that ARVD/C is a disorder of the desmosome and provided insight into its pathogenesis [2-5]. Desmosomal impairment followed by mechanical and electric uncoupling of cardiomyocytes leads to cell death with fibrofatty replacement [6]. Two nondesmosomal genes have been associated with specific types of ARVD/C: patients with CPVT have mutations in the cardiac Ryanodine receptor gene [7].

The prevalence of ARVD/C is unknown, it is thought to occur in six per 10,000 persons in certain populations. Some studies have suggested that it may be as common as 1/1,000, reaching up to 1/200 in carriers of mutations relevant to ARVD/C, accounting for up to 17% of all sudden cardiac deaths in the young [8]. After hypertrophic heart disease, it is the number one cause of sudden cardiac death in young persons, especially athletes. The initial diagnosis of ARVD/C is based on the criteria established in 1994 [7] and revised in 2010 [9] including pathogenic mutations as a diagnostic criteria. After the diagnosis is made in one person, all of his or her first-degree relatives should be screened. Treatment focuses on controlling the arrhythmias and in managing any signs or symptoms of heart failure. Antiarrhythmic medications are the initial and most commonly used therapy in ARVD/C. No single drug has been shown to be completely effective in controlling the arrhythmias. The therapeutic options are limited like pharmacological agents (ACEI, anticoagulants, diuretics, and antiarrhythmic agents as sotalol, verapamil, betablockers, amiodarone, and flecainide), catheter ablation to eliminate drug resistant conduction pathways (critical to the perpetuation of arrhythmias), implantable defibrillators in refractory patients at risk for sudden death and surgery as the last option, consisting on ventriculotomy and disconnection of the RV free wall or cardiac transplantation if severe terminal heart failure [1, 6, 10, 11].

Thus, although ARVD/C is a clinically well defined disease, the molecular events underlying the disease phenotype have not reached an equal level of maturity. The specific aim of this paper is to understand the molecular forces behind the disease, and state them within a general molecular perspective.

There are essentially three forms, RyR1, RyR2 and RyR3. RyR1 is the channel in the skeletal muscle, RyR2 is the type expressed in the heart muscle, and RyR3 is found predominantly in the brain [12]. $Ca^{++}$ release from the sarcoplasmic reticulum, SR, mediated by the cardiac Ryanodine receptor (RyR2) is a fundamental event in cardiac muscle contraction. These receptors, which mediate the release of calcium from the SR to the cytosol, form a group of four homotetramers, with a large cytoplasmic assembly and a transmembrane domain called the pore region[13]. The first 3089 residues form the cytoplasmic region. The remaining 1777 residues, mostly helical in structure, form the transmembrane domain. The three dimensional structure of the full assembly is not known with high precision so as to merit structure based ligand design calculations. Effective computational methods are needed to characterize the various compartments of the assembly as effective targets [14]. However, the crystal structure of the first 173 amino acids of the N-terminal domain of the wild type RyR2 and its mutated form are determined with high precision by Petegem and Lobo [15] which are adopted in the present paper for calculations.

In the diseased system, calcium leakage is the most important factor that can generate delayed after depolarizations, which can lead to fatal arrhythmias. More than 300 point mutations have been identified in the Ryanodine receptor some of which are associated with the disorders observed clinically [8, 16-18]. The present paper deals with one of these disease-causing point mutations, the A77V mutation. However, before analyzing the structural consequences of the mutation, we carry out a detailed analysis of the structural features of RyR2. We use a simple elastic net model [19] which is described in some detail in the Supplementary section. This model is based on the identification of residues that are responsive to conformational energy fluctuations of the protein. The energy responsive residues lie on a pathway which is regarded as the energy conduction pathway in the protein along which information is transmitted through correlated conformational fluctuations. We show here that the evolutionarily conserved residues lie on this path, thus pointing to the possible relation between conserved and energetically responsive amino acids. Identification of an energy conduction path in the protein establishes a reference structure. We show that the A77V mutation disrupts this pathway, thereby obliterating means of transferring information through the protein.

**Methods**: The N-terminal domain of RyR2 is a signal protein of 173 amino acids. The crystal structure of the N-terminal domain of physiological RyR2 (PDB code 3IM5) and the A77V mutated crystal structure (PDB code 3IM7) have been determined by x-ray with resolutions of 2.5 and 2.2 Å, respectively, by Van Petegem and Lobo [15]. The protein consists of a β-trefoil of 12 β strands held together by hydrophobic forces. A 10-residue α helix is packed against strands β4 and β5. A 3 residue 3-10 helix is present in the loop containing β3 and β4. The N-terminal contains two Mir domains, similar to inositol 1,4,5-triphosphate receptor (IP3R), which is a more widely studied receptor whose function is to regulate the amplitude and frequency of $Ca^{++}$ oscillations.

*Determining the energy conduction path:* Elastic net models are simple, $C^\alpha$ based coarse grained models which can identify residues that play important role on the energetic interactions of the protein. A detailed explanation of the model that identifies the residues taking part in energy transfer is given in the Supplementary section. Stated briefly, the picture is as follows: Fluctuations $\Delta R_i$ in the position of a residue i, resulting from thermal energy, cause the atoms of the residue to interact with the atoms of the neighboring residues. On the average, a residue has about twelve neighboring residues that it directly interacts with. These fluctuation based interactions induce changes in the energy of the residue and its neighbors because the distance between each interacting ith and jth residues changes. The resulting fluctuation of the energy over the distance $R_{ij}$ is $\Delta \hat{U}_{ij}$. The hat over $U$ denotes that this is an instantaneous value. The energetic interaction of residue i with all other residues of the protein is expressed by the sum $\Delta \hat{U}_i = \sum_j \Delta \hat{U}_{ij}$. The energy response $\Delta \hat{U}_i$ of i may be correlated with the energy response $\Delta \hat{U}_j$ of residue j. This correlation is given with the nonzero value of the average $\langle \Delta \hat{U}_i \Delta \hat{U}_j \rangle$ where the angular brackets denote an average over all instantaneous fluctuations. Summing up over the fluctuations of the jth residue leads to the energy response $\Delta U_i$ of residue i as $\Delta U_i = \sum_j \langle \Delta \hat{U}_i \Delta \hat{U}_j \rangle$. This is a thermodynamically meaningful quantity showing the mean energy response of residue i to all fluctuations of its surroundings. These correlations extend throughout the protein, leading to specific paths along which the fluctuations propagate. Recent work shows that these paths are evolutionarily conserved [20]. The simplest elastic net model assumes harmonic fluctuations of the residues and the energies [19]. More sophisticated versions of elastic networks introduce anharmonicities into the model for more accurate results at the expense of simplicity. We apply the harmonic model to the analysis of wild type RyR2 and its A77V mutated form.

*The energy conduction path of RyR2*: The residues that yield high values of the energy response are calculated according to the scheme outlined in the Supplementary section. In Figure 1, the mean energy response $\langle \Delta U_i \rangle$ of residue i is presented along the ordinate as a function of residue index. The circles indicate the conserved residues of 3IM5, obtained from the work of Goldenberg et al [21] (See also the PDBSum web site [22]). Comparison of the solid curve peaks and the circles shows that there is a strong correlation between the energy conduction path residues and conserved residues, which is in agreement with the recent suggestion of Lockless and Ranganathan [20]. The set of conserved residues of the protein all lie within the set of energetically important residues and are located along or in the neighborhood of energy conduction path. The peaks of Figure 1 constitute a path of residues that are neighbors in space. These peaks correspond to pairs of neighboring residues that exhibit large correlation values $\langle \Delta U_i \Delta U_j \rangle$ of energy responsive residues i and j. The method of determining the correlated residues is given in the Supplementary section. In three dimensions, these residues are shown in Figure 2a in yellow highlight. Figure 2b is obtained by 90° rotation around the vertical axis. For a clearer representation, the path residues are shown and identified in Figure 3. It is clearly seen that the path consists of residues that are hydrogen bonded to each other, as shown by the green dashed lines.

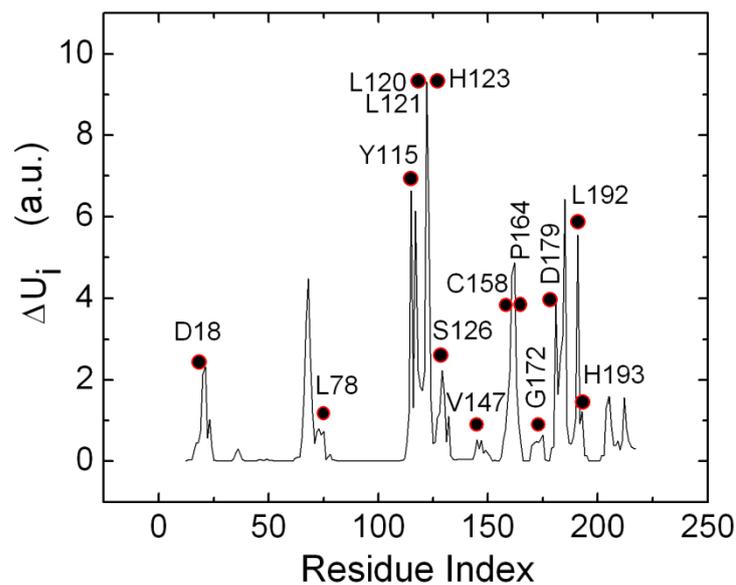

Figure 1. Energetically responsive residues (solid line) obtained with the elastic net model described in the Supplementary section, and the conserved residues (circles) obtained from Reference [21]. The ordinate values are in arbitrary un-normalized units.

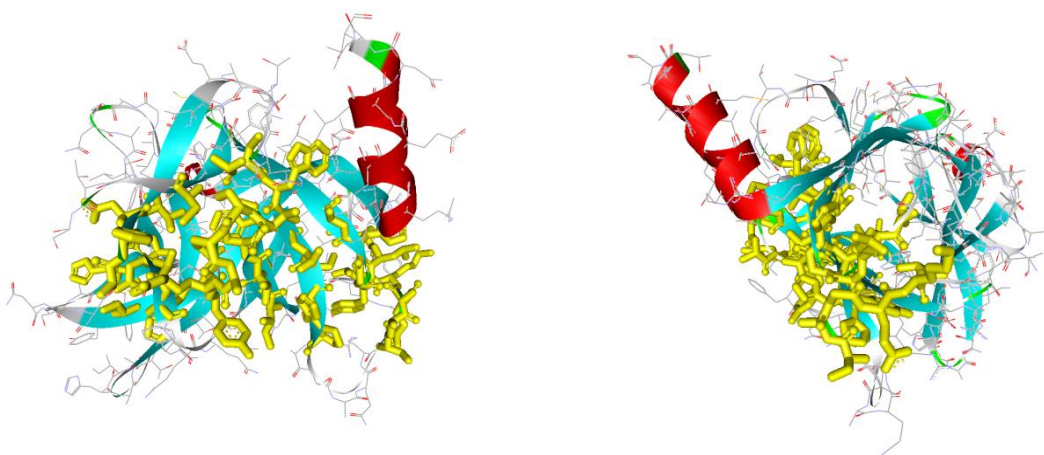

Figure 2a. Residues, highlighted in yellow, along the energy conduction path in RyR2. Figure 2b. 90° rotated depiction of Figure 2a.

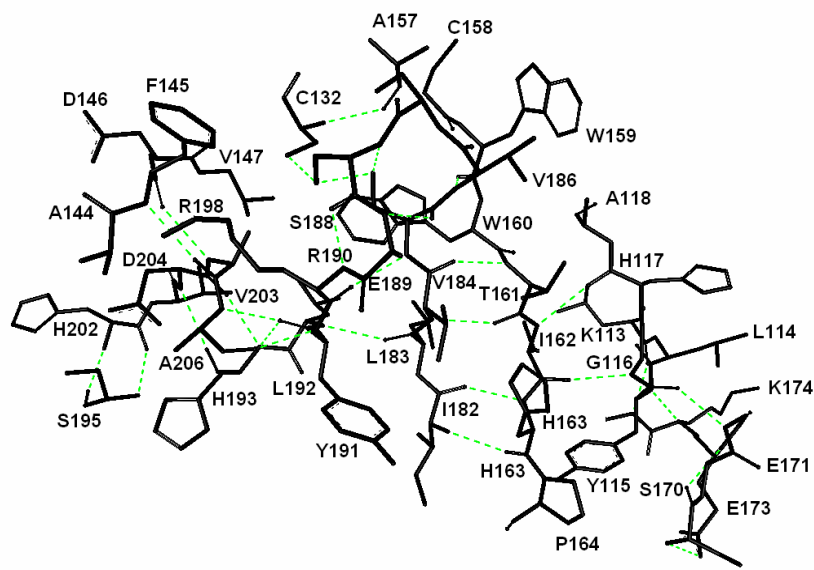

Figure 3. Residues along the energy conduction pathway. Green dashed lines denote the hydrogen bonds.

RyR2 contains two interspersed Mir domains, residues 124-178 and 164-217 [23]. Elastic net calculations show that the residues that lie on the energy conduction pathway are closely associated with the Mir domains. In Figure 4, we show the Mir domains in solid ribbon and the energetically responsive residues by black lines. The green dashed lines show the hydrogen bonds. The energetically responsive residues either lie on the Mir domains, or they are hydrogen bonded to these domains. There is no residue that is not closely associated with the Mir domain. We conclude that the

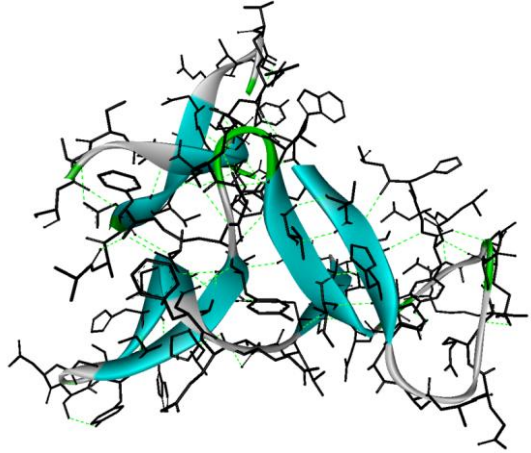

Figure 4. The two Mir domains shown in solid ribbon and the energetically responsive residues shown in black lines.

Mir domains of RyR2 plays an active role in energy transport through the protein. Below, we arrive at the same conclusion for the similar protein, the IP3R.

*Effects of A77V mutation*: The mutation A77V in RyR2 introduces two extra methyl groups, which induces rearrangements in the neighborhood of the mutation. However, these rearrangements appear to be correlated with the rest of the structure because there are large differences in the equilibrium conformations of the two structures as may be verified from the difference map of Figure 5. The extent of conformational changes induced by the mutation may be understood by comparing the corresponding residue pair distances $R_{ij} = |\bm{R}_j - \bm{R}_i|$, where $\bm{R}_i$ is the position vector of the ith $C^\alpha$.

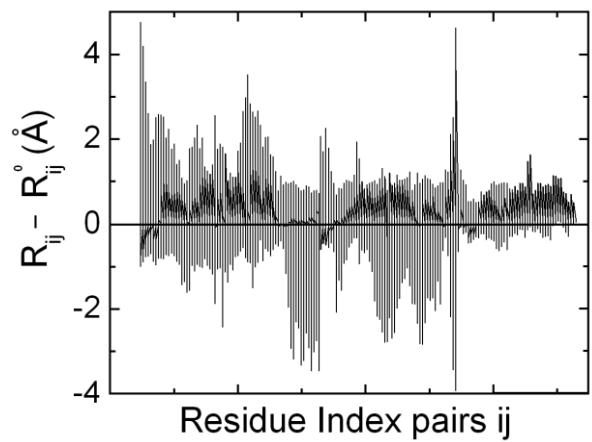

*Figure 5. The differences between $R_{ij} - R_{ij}^{\,0}$ of the mutated and wild type $C^\alpha$ separations.*

In Figure 5, we plot the quantity $R_{ij} - R_{ij}^0$ for the two structures, 3IM5 and 3IM7, as a function of residue pair indices ij, where the superscript zero denotes the wild type. The abscissa contains n(n+1)/2 = 15051 points corresponding to the residue pairs. Most of the distance differences are significant and above the uncertainties of the experimental resolution of 2.2-2.5 Å level. As will be shown below, the effects induced by these differences are not only local and confined to the vicinity of the mutation. The mutation introduces effects that are correlated with the overall protein structure, which in turn introduce modifications in the energy conduction pathway.

Elastic net calculations similar to the wild type are performed on the A77V mutated RyR2 in order to elucidate the differences between the energetically responsive residues. Results show that the energy conduction path observed in the wild type is disrupted in the mutated protein. The correlation data, $\langle \Delta U_i \Delta U_j \rangle$, indicate that the residues shown in red in Figure 6 are missing along the enegy conduction path of the mutated protein. The identities of the missing residues are indicated in the figure.

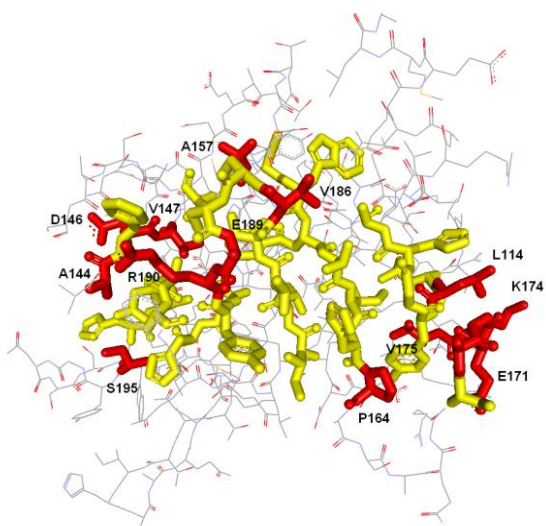

Figure 6. The residues that loose correlation upon mutation, shown in red and identified by residue type and number.

*Insights from the Inositol Receptor:* The relationship between the structure and function of RyR2 is not well understood. Here, we apply the elastic net analysis to IP3R (PDB code 1N4K) in order to gain insight into the less known RyR2. IP3R, similar to RyR2 is a $Ca^{2+}$ release channel, with structural similarity to RyR2. Both have a trefoil region each containing Mir domains. In Figure 7, we present the energetically responsive residues (solid lines) and the conserved residues (circles) obtained by the elastic net analysis. Similar to RyR2, the conserved residues lie either on the energy conduction path or in its immediate neighborhood. In Figure 8, we present the three dimensional picture of the protein, with the calculated energy correlation path shown in thick black lines. On the left of the figure, the green colored molecule is IP3. It makes one bond with R265 of the energy correlation path. On the right hand side, the two glutamic acid residues form the Calcium binding site. Similar to RyR2, the energetically responsive residues are located either on the Mir domain, or are hydrogen bonded to it.
6

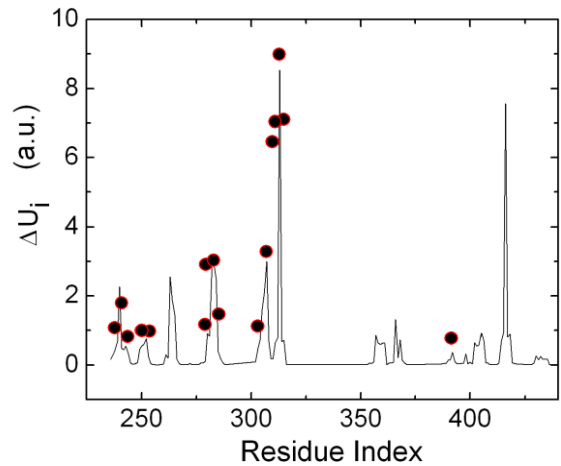

Figure 7. Energetically responsive residues (solid line) and the conserved residues (circles) of the trefoil part of 1N4K. See the legend for Figure 1.

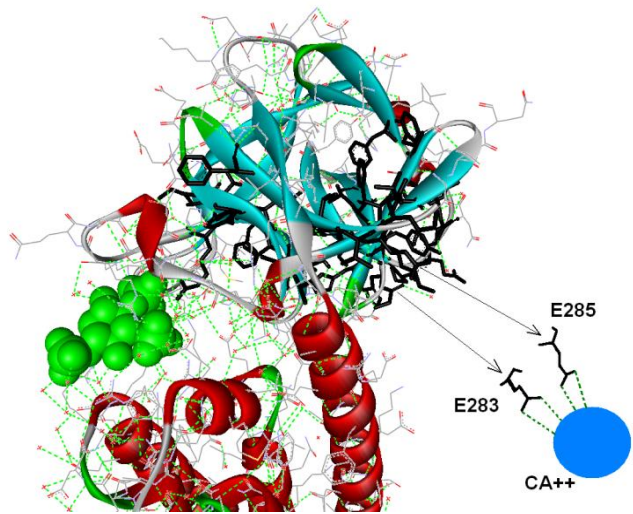

Figure 8. Energetically responsive residues of IP3R highlighted in black. The green colored molecule is IP3. On the right of the path, the two residues E283 and E285 constitute the calcium binding site. The offset shows the binding mode of calcium to the two residues, the green dashed lines are the hydrogen bonds.

**Discussion**: Using a simple computational model, we identified an energy conduction pathway for the wild type RyR2. This path whose residues are shown in Figure 3 has several features of interest. Firstly, it contains most of the evolutionarily conserved residues. The remaining conserved residues are in the close neighborhood of the path, all making hydrogen bonds with the residues of the path. This important feature has recently been shown by Lockless and Ranganathan [20] implying that evolutionary conservation is driven by energy conduction in proteins. Although no ligands for RyR2 have been observed until now [24], the two glutamic acids, E171 and E173 on the extremity of the path may potentially form a calcium binding site. This suggestion is based on the observation that in IP3R a known calcium binding site is formed by E283 and E285 whose location on the path coincides exactly with that of RyR2. That the residue E173 (E164 in RyR1) is exposed to water and is a candidate for possible binding is also suggested by Hamilton [14]. The correspondence of the glutamic acid sites for the two proteins, RyR2 and IP3R can be seen by comparing Figures 3 and 8. Several residues of the energy conduction path of the wild type RyR2 dissapear on the path of the disease causing A77V mutation. Among the dissapearing residues is E171, as well as several others on the two extremities of the path as identified and shown in red in Figure 6. The loss of energy response of these residues is expected to disrupt signal conduction that would otherwise result from the correlated fluctuations of the residues along the path. These suggestions are based on the elastic net analysis of the more widely studied calcium channel receptor IP3R.

**Acknowledgment**: We are grateful for the suggestions of Professor Filip van Petegem for insightful suggestions.

References

1. Patel, H.C. and H. Calkins, *Arrhythmogenic right ventricular dysplasia.* Current treatment options in cardiovascular medicine, 2010: p. 598-613.
2. Gerull, B., et al., *Mutations in the desmosomal protein plakophilin-2 are common in arrhythmogenic right ventricular cardiomyopathy (vol 36, pg 1162, 2004).* Nature Genetics, 2005. **37**(1): p. 106-106.
3. Pilichou, K., et al., *Mutations in desmoglein-2 gene are associated to arrhythmogenic right ventricular cardiomyopathy.* European Heart Journal, 2006. **27**: p. 294-294.
4. McKenna, W.J., S. Sen-Chowdhry, and P. Syrris, *Genetics of right ventricular cardiomyopathy.* Journal of Cardiovascular Electrophysiology, 2005. **16**(8): p. 927-935.
5. McKenna, W.J., et al., *Arrhythmogenic right ventricular dysplasia/cardiomyopathy associated with mutations in the desmosomal gene desmocollin-2.* American Journal of Human Genetics, 2006. **79**(5): p. 978-984.
6. Saffitz, J.E., et al., *Structural and molecular pathology of the heart in Carvajal syndrome.* Cardiovascular Pathology, 2004. **13**(1): p. 26-32.
7. Abraham, T.P., et al., *Comparison of novel echocardiographic parameters of right ventricular function with ejection fraction by cardiac magnetic resonance.* Journal of the American Society of Echocardiography, 2007. **20**(9): p. 1058-1064.
8. Marks, A.R. and M.J. Betzenhauser, *Ryanodine receptor channelopathies.* Pflugers Archiv-European Journal of Physiology, 2010. **460**(2): p. 467-480.
9. Cox, M.G.P.J., et al., *Arrhythmogenic Right Ventricular Dysplasia/Cardiomyopathy Diagnostic Task Force Criteria Impact of New Task Force Criteria.* Circulation-Arrhythmia and Electrophysiology, 2010. **3**(2): p. 126-133.
10. Moric-Janiszewska, E. and G. Markiewicz-Loskot, *Review on the genetics of arrhythmogenic right ventricular dysplasia.* Europace, 2007. **9**(5): p. 259-266.
11. Danieli, G.A., et al., *Identification of mutations in the cardiac ryanodine receptor gene in families affected with arrhythmogenic right ventricular cardiomyopathy type 2 (ARVD2).* Human Molecular Genetics, 2001. **10**(3): p. 189-194.
12. Kimlicka, L. and F. Van Petegem, *The structural biology of ryanodine receptors.* Science China-Life Sciences, 2011. **54**(8): p. 712-724.
13. http://www.uniprot.org/uniprot/Q92736. [cited 11, 09, 2011].
14. Hamilton, S.L. and I.I. Serysheva, *Ryanodine Receptor Structure: Progress and Challenges.* Journal of Biological Chemistry, 2009. **284**(7): p. 4047-4051.
15. Van Petegem, F. and P.A. Lobo, *Crystal Structures of the N-Terminal Domains of Cardiac and Skeletal Muscle Ryanodine Receptors: Insights into Disease Mutations.* Structure, 2009. **17**(11): p. 1505-1514.
16. Van Petegem, F. and L. Kimlicka, *The structural biology of ryanodine receptors.* Science China-Life Sciences, 2011. **54**(8): p. 712-724.
17. Priori, S.G. and S.R.W. Chen, *Inherited Dysfunction of Sarcoplasmic Reticulum Ca(2+) Handling and Arrhythmogenesis.* Circulation Research, 2011. **108**(7): p. 871-883.
18. Williams, A.J., et al., *Ryanodine receptor mutations in arrhythmia: The continuing mystery of channel dysfunction.* Febs Letters, 2010. **584**(10): p. 2153-2160.
19. Erman, B., *Relationships between ligand binding sites, protein architecture and correlated paths of energy and conformational fluctuations.* Physical Biology, 2011. **8**: p. 056003 (9pp).

20. Lockless, S.W. and R. Ranganathan, *Evolutionarily conserved pathways of energetic connectivity in protein families.* Science, 1999. **286**(5438): p. 295-299.
21. Goldenberg O., et al., *The ConSurf-DB: Pre-calculated evolutionary conservation profiles of protein structures.* Nucleic Acids Research, 2009. **37**: p. Database issue D323-D327.
22. PDBSUM. *http://www.ebi.ac.uk/pdbsum/*.   [cited.
23. Amador, F.J., et al., *Crystal structure of type 1 ryanodine receptor amino-terminal beta-trefoil domain reveals a disease-associated mutation "hot spot" loop.* Proceedings of the National Academy of Sciences of the United States of America, 2009. **106**(27): p. 11040-11044.
24. van Petegem, F., *Private correspondence* .

Supplementary Information on the Elastic Net Model

Proteins perform their function through the correlated fluctuations of their amino acids. Although all atom molecular dynamics simulations are capable of yielding valuable information on the functional motions of proteins, only very short times can be reached which is too short for a complete picture of protein behavior. Therefore, coarse grained models that can give an approximate but a more complete and overall picture of protein behavior are needed. Elastic net models are suitable for this purpose. The Gaussian Network Model (GNM) is a simple coarse grained elastic net model that predicts the fluctuations of amino acids and the correlations between these fluctuations. Detailed information on GNM can be found in Reference [1]. The only input required for the analysis is the three dimensional structure of the protein. Coarse graining refers to eliminating the details of the atomic structure by keeping only a few representative atoms. In the GNM, only the alpha carbons, $C^\alpha$'s are considered. All other atoms are assumed to be collapsed on the respective $C^\alpha$ of the chain. We refer to an amino acid represented by the $C^\alpha$ as the residue. Each residue is covalently attached to its neighboring residue along the primary structure of the chain. In addition to the two covalent neighbors of a residue, there are several spatial neighbors. Each residue interacts with these spatially neighboring residues, and in its simplest form, this interaction may be visualized as if these two residues are joined by a spring. The spring is representative of the more complex forces between the neighboring residues, such as hydrogen bonding, covalent bonding, electrical forces, etc. If two neighboring residues tend to separate from each other, the spring pulls them together and if they tend to get closer, the spring pushes them apart. In this way, the two residues interact with each other. In the GNM the spring is assumed to be linear, which obeys the relation $\Delta F = \gamma \Delta R$, where $\Delta R$ is the change in the distance between the two residues, $\Delta F$ is the force that arises from this change in distance, and $\gamma$ is the spring constant that reflects the stiffness of the interaction. The spring that is expressed by the linear force displacement law shows similar behavior in tension and compression, and the model defined by a linear spring is called a harmonic model. The GNM is based on harmonic interactions, but in reality, the forces involved between neighboring residues are not harmonic. In the coarse graining of the GNM, the spring constant between all pairs of neighboring residues, including the covalent ones, is taken to be of equal magnitude. Two residues are assumed neighbors in space if they are closer to each other than a given cutoff distance. This distance corresponds to the radius of the first coordination shell around a given residue, and is usually taken to be between 6.5Å-7.0Å. The knowledge of the three dimensional structure of the protein that has n residues allows us to write a connectivity matrix, C, where the rows and the columns identify the residue indices, from 1 to n, where the amino end is the starting and the carboxy end is the terminating end of the protein. If two residues i and j are within the cutoff distance, then $C_{ij} = 1$, otherwise it is zero. The relationship of the forces to the displacements is given by the equation $\Delta F_i = \sum_j \Gamma_{ij} \Delta R_j$ where $\Gamma_{ij}$ is an nxn matrix and is called the force constant matrix which we define below. $\Delta F_i$ is a vector of n entries, where i denotes the residue index and goes from 1 to n. Similarly, $\Delta R_j$ is a vector of n entries, where j goes from 1 to n. In words, this equation says that the force, $\Delta F_i$, acting on the ith residue is the linear combination of the contributions from the fluctuations, $\Delta R_j$, of all the neighboring residues where j goes from 1 to n. The matrix $\Gamma_{ij}$ is simply related to the connectivity matrix $C_{ij}$ by

$$\Gamma_{ij} = \begin{cases} -\gamma C_{ij} & \text{if } i \neq j \\ -\sum_k C_{ik} & \text{if } i = j \end{cases} \quad (S1)$$

The matrix $\Gamma_{ij}$ reflects the role of protein structure on fluctuations. Techniques of statistical mechanics allow us to derive several relationships between the fluctuations of residues [2-4]. The correlation between the fluctuations of residues i and j is related, for example, to the inverse of the matrix $\Gamma_{ij}$. The inverse of an nxn matrix is also an nxn matrix, which can easily be evaluated by freely available computer software. The relationship is

$$\langle \Delta R_i \Delta R_j \rangle = k_B T \, \Gamma^{-1}{}_{ij} \quad (S2)$$

Here, the angular brackets denotes the time average of the product of fluctuations of residues i and j, $k_B$ is the Boltzmann constant, $T$ is the physiological temperature expressed in Kelvin scale, $\Gamma^{-1}$ is the inverse matrix of $\Gamma$, and its subscripts i and j acknowledge the residue indices of interest. If i =j, then Eq S2 becomes

$$\langle \Delta R_i^2 \rangle = k_B T \, \Gamma^{-1}{}_{ii} \quad (S3)$$

The left hand side of Eq, S3 is the mean-square fluctuations of the ith residue which is related to experimentally available B-factors, $B_i$, through the equation

$$\langle \Delta R_i^2 \rangle = \frac{3}{8\pi^2} B_i \quad (S4)$$

The mean-square fluctuations $\langle \Delta R_{ij}^2 \rangle$ of the distance $\Delta R_{ij}$ between residues i and j are obtained from Eq. S3 as

$$\langle \Delta R_{ij}^2 \rangle = \Gamma^{-1}{}_{ii} - 2\,\Gamma^{-1}{}_{ij} + \Gamma^{-1}{}_{jj} \quad (S5)$$

The derivation of Eq S5 is given in Reference [5]. One can see from the foregoing explanations that several factors contribute to the fluctuations of residues. The distance between residues i and j are under various effects coming from all other residue fluctuations. The matrix $\Gamma^{-1}{}_{ij}$ contains several such effects coming from the structure of the protein.

Large scale fluctuations of the protein and local fluctuations can be separated from each other by a technique called mode analysis. Each mode represents coordinated motions of the protein. The chaotic-like motions of the protein are obtained by the superposition of the coordinated motions represented by the modes. Each mode is expressed by an eigenvector and the corresponding eigenvalue. Modes that correspond to the large eigenvalues of the $\Gamma$ matrix show local events and small eigenvalues correspond to global motion effects. We are interested in the residue based local events, therefore in the large eigenvalues and their corresponding eigenvectors. The largest few eigenvalues usually suffice to reflect the local effects on correlations of fluctuations. Here, we superpose the effects of largest five eigenvalues, the necessity and the sufficiency of which is explained in Reference [6]. Statistical mechanics gives the extent of correlation of one residue with the others in the protein as may be seen in more detail in Reference [6]. The magnitude $\Delta \hat{U}_{ij}$ of the fluctuation of energy resulting from the change in the distance between residues i and j is given for the harmonic system as $\Delta \hat{U}_{ij} = \frac{1}{2}\gamma C_{ij} \left| \Delta \hat{R}_j - \Delta \hat{R}_i \right|^2$ where, $\Delta \hat{R}_i$ is the instantaneous fluctuation of the ith residue from its native position. If we focus on residue i and superpose all of its energetic interactions with all of its neighbors, we obtain a measure of the energy response of i. This is expressed by the sum: [6]

$$\Delta \hat{U}_i = \sum_j \Delta \hat{U}_{ij} = \frac{1}{2}\gamma \sum_j C_{ij} \left(\Delta \hat{R}_j - \Delta \hat{R}_i\right)^2 \quad \text{(S6)}$$

$\Delta \hat{U}_i$ is a positive quantity, showing the energy response of residue i to the fluctuations of all other residues. The energy fluctuations of the residue i may influence the fluctuations of residue j. We that say that the two fluctuations are correlated. The correlation is given by [6]

$$\langle \Delta \hat{U}_i \Delta \hat{U}_j \rangle = \frac{1}{4}\gamma^2 \sum_j \sum_l C_{ij} C_{kl}$$
$$\times \left\langle \left(\Delta \hat{R}_i - \Delta \hat{R}_j\right)^2 \left(\Delta \hat{R}_k - \Delta \hat{R}_k\right)^2 \right\rangle \quad \text{(S7)}$$

The average shown by the angular brackets is given in terms of the elements of the $\Gamma$ matrix. If Equation S7 is summed over all residues j of the protein, we obtain the average value $\Delta U_i$

$$\Delta U_i = \sum_j \langle \Delta \hat{U}_i \Delta \hat{U}_j \rangle \quad \text{(S8)}$$

Physically, $\Delta U_i$ is an indication of the extent of energetic interactions of residue i with those of all other residues of the protein. We term it as the energy correlation of residue i with the rest of the protein, or simply the energy response of residue i. The expressions $\langle \Delta \hat{U}_i \Delta \hat{U}_j \rangle$ and $\Delta U_i$ are the two main measures used in the text. In order to reduce cross referencing, we write below the expression for $\langle \Delta \hat{U}_i \Delta \hat{U}_j \rangle$:

$$\langle \Delta \hat{U}_i \Delta \hat{U}_k \rangle = \frac{1}{4}(kT)^2 \sum_j \sum_l C_{ij} C_{kl} \Big[ 2\left((\Gamma_{ik}^{-1})^2 + (\Gamma_{il}^{-1})^2 + (\Gamma_{jk}^{-1})^2 + (\Gamma_{jl}^{-1})^2\right)$$

$$+\Gamma_{ii}^{-1}\Gamma_{kk}^{-1} + \Gamma_{ii}^{-1}\Gamma_{ll}^{-1} + \Gamma_{jj}^{-1}\Gamma_{kk}^{-1} + \Gamma_{jj}^{-1}\Gamma_{ll}^{-1}$$

$$-4\left(\Gamma_{il}^{-1}\Gamma_{ik}^{-1} + \Gamma_{jl}^{-1}\Gamma_{jk}^{-1} + \Gamma_{ik}^{-1}\Gamma_{jk}^{-1} + 2\Gamma_{il}^{-1}\Gamma_{jl}^{-1}\right) \quad \text{(S9)}$$

$$-2\left(\Gamma_{ii}^{-1}\Gamma_{kl}^{-1} + \Gamma_{jj}^{-1}\Gamma_{kl}^{-1} + \Gamma_{kk}^{-1}\Gamma_{ij}^{-1} + \Gamma_{ll}^{-1}\Gamma_{ij}^{-1}\right)$$

$$+4\left(\Gamma_{ij}^{-1}\Gamma_{kl}^{-1} + \Gamma_{ik}^{-1}\Gamma_{jl}^{-1} + \Gamma_{il}^{-1}\Gamma_{kj}^{-1}\right) \Big]$$

**References**
1. http://en.wikipedia.org/wiki/Gaussian_network_model. [cited.
2. Haliloglu, T. and B. Erman, *Analysis of Correlations between Energy and Residue Fluctuations in Native Proteins and Determination of Specific Sites for Binding* Physical Review Letters, 2009. **102**: p. 088103-088106.
3. Yogurtcu, O.N., M. Gur, and B. Erman, *Statistical thermodynamics of residue fluctuations in native proteins.* Journal of Chemical Physics, 2009. **130**(9): p. 095103-13.

4. Haliloglu, T., A. Gul, and B. Erman, *Predicting Important Residues and Interaction Pathways in Proteins Using Gaussian Network Model: Binding and Stability of HLA Proteins.* Plos Computational Biology, 2010. **6**(7): p. e1000845.
5. Tuzmen, C. and B. Erman, *Identification of Ligand Binding Sites of Proteins Using the Gaussian Network Model.* Plos One, 2011. **6**(1): p. e16474.
6. Erman, B., *Relationships between ligand binding sites, protein architecture and correlated paths of energy and conformational fluctuations.* Physical Biology, 2011. **8**: p. 056003 (9pp).